\documentclass[twocolumn,amsmath,amssymb,showpacs,superscriptaddress,pra]{revtex4}

\usepackage{graphicx}
\usepackage{dcolumn}
\usepackage{bm}
\usepackage{xcolor}

\newcommand{\bs}{\boldsymbol}
\newcommand{\mb}{\mathbf}

\begin{document}


\title{Wigner rotations and an apparent paradox in relativistic quantum information}

\author{Pablo L. Saldanha}\email{saldanha@fisica.ufmg.br}
\affiliation{Department of Physics, University of Oxford, Clarendon Laboratory, Oxford, OX1 3PU, United Kingdom}
\affiliation{Departamento de F\'isica, Universidade Federal de Pernambuco, 50670-901, Recife, PE, Brazil}
\affiliation{Departamento de F\'isica, Universidade Federal de Minas Gerais, Caixa Postal 702, 30161-970, Belo Horizonte, MG, Brazil}
\author{Vlatko Vedral}
\affiliation{Department of Physics, University of Oxford, Clarendon Laboratory, Oxford, OX1 3PU, United Kingdom}
\affiliation{Centre for Quantum Technologies, National University of Singapore, Singapore}
\affiliation{Department of Physics, National University of Singapore, Singapore}

\date{\today}

\begin{abstract} 
It is shown that a general model for particle detection in combination with a linear application of the Wigner rotations, which correspond to momentum-dependent changes of the particle spin under Lorentz transformations, to the state of a massive relativistic particle in a superposition of two counter-propagating momentum states leads to a paradox. The paradoxical behavior is that the probability of finding the particle at different positions would depend on the reference frame. A solution to the paradox is given when the physical construction of the corresponding state is taken into account, suggesting that we cannot in general linearly apply the Wigner rotations to a quantum state without considering the appropriate physical interpretation.   
\end{abstract}

   
                           
\maketitle

\section{Introduction}

A Wigner rotation corresponds to a momentum-dependent change of the spin state of a relativistic particle with a change of reference frame \cite{wigner49,weinberg}. It is a direct consequence of the imposition of the special relativistic space-time structure to quantum mechanics. It is thus deeply connected with the basic structure of the universe that, as far as we know, is both quantum and relativistic. The influence of the Wigner rotations on the field of quantum information has been intensively investigated in the past 10 years \cite{peres02,alsing02,gingrich02,ahn03,terashima03,li03,peres04,lee04,li04,gonera04,bartlett05,kim05,czachor05,peres05,caban05,jordan06,lamata06,jordan07,landulfo09,dunningham09,friis10,choi11,esfahani12,palge12,saldanha12a,saldanha12b}. Since the seminal paper of Peres, Scudo and Terno \cite{peres02}, who concluded that because of the momentum-dependent Wigner rotations the spin entropy of a relativistic particle is not a relativistic scalar, many works appeared in the literature discussing how the entropy \cite{peres02,peres04,li04,gonera04,bartlett05,peres05,caban05,jordan06,dunningham09,palge12} and entanglement \cite{gingrich02,li03,lamata06,jordan07,friis10,choi11,esfahani12} of the reduced spin state of a relativistic system change under Lorentz transformations, as well as the influence of the Wigner rotations on the violation of Bell's inequalities with relativistic particles \cite{ahn03,terashima03,lee04,kim05,landulfo09,saldanha12b}. However, in our previous works \cite{saldanha12a,saldanha12b} we showed that it is not possible to consistently define a reduced spin density matrix for a system with one or more relativistic particles, as it is done in most of the cited papers  \cite{peres02,gingrich02,li03,peres04,li04,gonera04,bartlett05,peres05,caban05,jordan06,lamata06,jordan07,landulfo09,dunningham09,friis10,choi11,esfahani12,palge12}. 

Here we go  further. We show that if, under a change of the reference frame, we simply linearly apply Wigner rotations to the quantum state of a massive particle that is in a superposition of two counter-propagating momentum states and consider a general model for particle detection, we obtain a paradox, since the probability of finding the particle in a given position would depend on the reference frame. A solution of the paradox is given here based on the physical interpretation of the Wigner rotations recently given by us \cite{saldanha12a} and on a discussion about the preparation method of the quantum state of the particle. In particular, we show that the Wigner rotation depends on how the particle's quantum state is prepared, such that it is not possible to compute the rotation for each momentum component separately, a subtle consideration that removes the paradox. In other words, the solution we present for the paradox  is based on the fact that the Wigner rotation operation cannot in general be linearly applied to an arbitrary superposition of different momentum states. Our conclusions affect much of the literature on relativistic quantum information which has to be re-evaluated in order to avoid  inconsistencies like the one to be presented here.

\section{Physical system of the paradox}

Consider the case of a spin-1/2 massive relativistic particle in the quantum state 
\begin{equation}\label{psi}
	|\Psi\rangle=\frac{1}{\sqrt{2}}\Big[ |p\mb{\hat{y}},+Z\rangle - |-p\mb{\hat{y}},+Z\rangle \Big]
\end{equation}
in reference frame $S^{(0)}$, where $|\mb{p},\pm Z\rangle$ represents a state for the particle with 4-momentum $(p^0,\mb{p})$, with $p^0=\sqrt{m^2c^4+c^2|\mb{p}|^2}$, and spin state pointing in the $\pm\mb{\hat{z}}$ direction, being the eigenvector of the Pauli matrix $\hat{\sigma}_z$ with eigenvalue $\pm1$. 
We are using Wigner's definition for spin \cite{wigner49}, that refers to the particle angular momentum in the rest frame for each momentum component. From now on we will use a system of units in which the speed of light in vacuum is $c=1$. 

Using the Foldy-Wouthuysen transformation on the Dirac Hamiltonian for a spin-1/2 massive relativistic particle, the $z$ component of the mean spin operator is a constant of motion of the free Hamiltonian and the mean position operator is independent of spin \cite{foldy50}. It is in the Foldy-Wouthuysen representation that the Pauli matrix $\hat{\sigma}_z$ is the $z$ component of the mean spin operator of a relativistic particle as we consider in this work \cite{foldy50}. The components of the particle wavefunction with mean spin state $|\pm Z\rangle$ in the mean position representation can be written as \cite{grandy}
\begin{equation}\label{wavefunction}
	\Psi_{\pm Z}(\mb{r})=\int d^3p K(p^0) \mathrm{e}^{i{\mb{p}\cdot\mb{r}}/{\hbar}}\langle \mb{p},\pm Z | \Psi\rangle,
\end{equation}
where $K(p^0)$ is a factor that depends on the specific position operator that we use. Since the state of Eq. (\ref{psi}) have a superposition of two momenta with equal magnitudes, our results do not depend on the specific form of $K(p^0)$. For the state of Eq. (\ref{psi}) we have $\Psi_{+Z}(\mb{r})\propto\sin({py}/{\hbar})$ and $\Psi_{-Z}(\mb{r})=0$. So the probability density of finding the particle around position $y$ obeys
\begin{equation}\label{p0}
	P_0(y)\propto\sin^2\left(\frac{py}{\hbar}\right).
\end{equation}

If we make a change of reference frame to a frame $S^{(1)}$ that moves with velocity $\beta \mb{\hat{z}}$ in relation to $S^{(0)}$, each momentum component of the state (\ref{psi}) suffers a different spin transformation due to the dependence of the Wigner rotation with the particle momentum. The spin transformations are \cite{halpern}
\begin{equation}\label{wigner transf}
	\hat{R}(\beta \mb{\hat{z}},\pm p\mb{\hat{y}})=\cos\left(\frac{\varphi}{2}\right)\hat{\sigma}_0\pm i\sin\left(\frac{\varphi}{2}\right)\hat{\sigma}_x,
\end{equation}
where $\hat{\sigma}_0$ represents the identity and $\hat{\sigma}_x$ the $x$ Pauli matrix, with
\begin{equation}\label{wigner ang}
	\sin\left(\frac{\varphi}{2}\right)=\sqrt{\frac{(\gamma_p-1)(\gamma_\beta-1)}{2(1+\gamma_p\gamma_\beta)}},
\end{equation}
where $\gamma_\beta\equiv 1/\sqrt{1-\beta^2}$ and $\gamma_p\equiv\sqrt{m^2+p^2}/m=1/\sqrt{1-v^2}$ if $\mb{v}$ is the particle velocity corresponding to the momentum $\mb{p}$. The momentum state of the particle also changes with the change of reference frame, but the $y$ component remains the same. Here we concentrate on the $y$ dependence of the particle wavefunction, so we will not worry about the momentum in the $x$ or $z$ directions. Of course, the state of Eq. (\ref{psi}) must be seen as an approximation, since the wavefunction of Eq. (\ref{wavefunction}) must decay to zero with large $x$ and $z$, but we will simply consider that the wavefunction can be decomposed as $\Psi_{\pm Z}(\mb{r})=\psi_{\pm Z}(y)\xi_{\pm Z}(x,z)$ and concentrate our discussion on $\psi_{\pm Z}(y)$. According to Eqs. (\ref{psi}), (\ref{wavefunction}) and (\ref{wigner transf}), in the new frame we have 
\begin{eqnarray}\label{psil}\nonumber
	\psi_{+Z}'(y)&\propto& \cos\left(\frac{\varphi}{2}\right) \sin\left(\frac{py}{\hbar}\right),\\\psi_{-Z}'(y)&\propto& \sin\left(\frac{\varphi}{2}\right) \cos\left(\frac{py}{\hbar}\right).
\end{eqnarray} 
 The probability density of finding the particle around position $y$ in the new frame obeys
\begin{eqnarray}\label{p1}
	P_1(y)&\propto&|\psi_{+Z}'(y)|^2+|\psi_{-Z}'(y)|^2\\\nonumber
	&\propto& \cos^2\left(\frac{\varphi}{2}\right)\sin^2\left(\frac{py}{\hbar}\right)+\sin^2\left(\frac{\varphi}{2}\right)\cos^2\left(\frac{py}{\hbar}\right).
\end{eqnarray}

In Fig. 1 we plot the modulus squared of the particle wavefunction in reference frame $S^{(0)}$, given by Eq. (\ref{p0}), and in reference frame $S^{(1)}$, given by Eq. (\ref{p1})  for $\gamma_\beta=10$ and $\gamma_p=1.2$ in Eq. (\ref{wigner ang}). It can be seen a loss of visibility in the interference of the two momentum components that compose the particle wavefunction in the new frame. In reference frame $S^{(0)}$ there is no correlation between the particle momentum and spin, and the state of Eq. (\ref{psi}) generates a complete destructive interference at the position $y=0$, as can be seen by the probability density of Eq. (\ref{p0}). In the reference frame $S^{(1)}$, on the other hand, the linear application of the Wigner rotations generates correlation between spin and momentum in the state of Eqs. (\ref{psil}). By tracing out spin, the reduced momentum wavefucntion is not pure anymore, such that there is no complete destructive interference at the position $y=0$, as can be seen in the probability density of Eq. (\ref{p1}).  This is a paradox. The probability of finding the particle around some position cannot depend on the reference frame.

\begin{figure}\begin{center}
  \includegraphics[width=7cm]{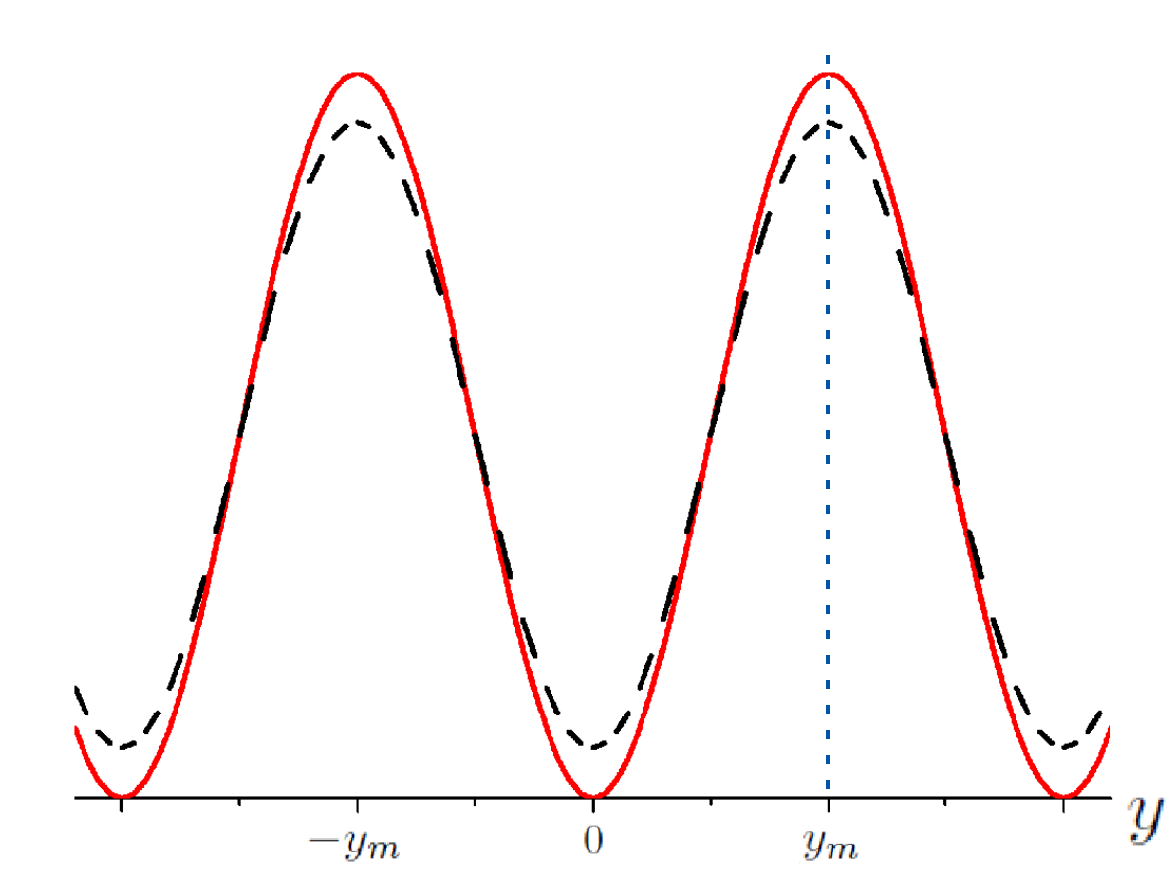}\\
  \caption{Modulus squared of the particle  wavefunction in reference frame $S^{(0)}$ (continuous red curve), given by Eq. (\ref{p0}), and modulus squared of the particle  wavefunction in reference frame $S^{(1)}$ (dashed black curve), given by Eq. (\ref{p1}) for $\gamma_\beta=10$ and $\gamma_p=1.2$ in Eq. (\ref{wigner ang}).} \label{fig1}
 \end{center}\end{figure}

Let us be more precise and consider measurements of the particle position using a detector that, by definition, responds only to the charge or the mass of the particle, but not to its spin. The probability of the particle detection with the central part of the detector placed at a position $y_c$ is assumed to be 
\begin{equation}\label{det}
	P(y_c)=\int dy \Gamma(y-y_c)|\psi(y)|^2,
\end{equation}
with $|\psi(y)|^2=|\psi_{+Z}(y)|^2+|\psi_{-Z}(y)|^2$. The exact form of $\Gamma(y)$ in (\ref{det}) depends on details of the detection scheme, but for the sake of simplicity we will consider $\Gamma(y)\propto\mathrm{e}^{-y^2/w^2}$. Since it is not possible to localize a particle in dimensions smaller than its Compton wavelength $\lambdabar=\hbar/(mc)$, we must have $w>\lambdabar$. We can define $R_0=P(0)/P(y_m)$ as the ratio between the probability of finding the particle around the minimum of the modulus of the wavefunction, at $y=0$, and the probability of finding it around the maximum, at $y=y_m$ in Fig. \ref{fig1} in reference frame $S^{(0)}$. Anagously, we define $R_1$as the same ratio in reference frame $S^{(1)}$. Considering that the superposition of momenta of the state (\ref{psi}) is not very relativistic, such that we can write $p\simeq mv$ and $\gamma_p\simeq 1+v^2/2$ up to the second order in $v$, it is straightforward to show, using Eqs. (\ref{det}), (\ref{p0}), (\ref{p1}), and (\ref{wigner ang}), that
\begin{equation}\label{R}
	R_0\simeq\frac{m^2w^2v^2}{2\hbar^2}\;,\;\frac{R_1}{R_0}\simeq1+\frac{(\gamma_\beta-1)}{2(\gamma_\beta+1)}\frac{\lambdabar^2}{w^2}.
\end{equation}
 For $\gamma_\beta\gg1$ and $w\simeq\lambdabar$, we have ${R_{\psi}}/{R_{\phi}}\approx 1.5$. Since the ratio between the probabilities in the two reference frames are different, we have a paradox. 

It is worth to discuss the relation of our calculations so far and Peres \textit{et al.} work \cite{peres02}.  In Ref. \cite{peres02}, the authors consider a pure state for a relativistic particle separable in the spin-momentum partition, thus having pure reduced states for spin and momentum in the considered reference frame. But the system may not be separable in another reference frame due to the momentum dependence of the Wigner rotations. The entanglement between spin and momentum in the new frame results in a mixed reduced spin density matrix in the new frame, such that the spin entropy is not a relativistic scalar \cite{peres02}. Here we are facing the same phenomenon for the state of Eq. (\ref{psi}), but considering the momentum reduced state. In the frame $S^{(1)}$ the reduced momentum state is not pure anymore due to the momentum-spin entanglement generated by the Wigner rotations, what causes the visibility reduction of the interference pattern of the position wavefunction represented in Fig. \ref{fig1}. It is important to reinforce that, as we showed in our previous work \cite{saldanha12a}, it is not possible to consistently define a reduced density matrix for the spin of a relativistic particle as done in Ref. \cite{peres02}, since it is not possible to measure the particle spin independently of its momentum in a relativistic setting. However, in principle it is possible to measure the momentum of a relativistic particle independently of its spin, such that the definition of a reduced momentum state should be reasonable.

\section{Solution of the paradox}

The deduction of the Wigner rotations always assumes free-particle states \cite{weinberg}. The state of Eq. (\ref{psi}) corresponds to the superposition of free-particle solutions, consequently being also a free-particle solution. However, how can one physically construct states like the one from Eq. (\ref{psi}), with a standing wave pattern? To obtain states like these, one must partially reflect the particle wavefunction, what is not possible with a uniform potential occupying the whole space. Although  Eq. (\ref{psi}) does represent a free-particle solution, the construction of such state needs the presence of a potential barrier, such that the simply application of the Wigner rotation to each momentum component is not a valid procedure (i.e. we are no longer in the domain of special relativity since the potential barrier accelerates the particle). 

To understand why this is the case, we can make use of the physical interpretation of the Wigner rotations recently given by us \cite{saldanha12a}, that says that these rotations are a consequence of the fact that different observers compute different quantization axes for a spin measurement.  This interpretation is supported by a recent work from Palmer \textit{et al.} that presents a detailed analysis of the Stern-Gerlach measurement process in a relativistic setting \cite{palmer12}. To compute the Wigner rotation for the state of  Eq. (\ref{psi}) with the change of reference frame, we must describe how the spin measurement is made and consider how the moving observer describes the quantization axis of this measurement. To construct the state, one can make a spin measurement on a particle that propagates in the $+\bs{\hat{y}}$ direction with a Stern-Gerlach apparatus with magnetic field in the $+\bs{\hat{z}}$ direction, obtaining eigenvalue $+1$. After that, the particle is sent to a region that has a potential barrier that reflects its wavefunction and produces the standing wave pattern. In this case, the moving observer will describe the Wigner rotation for both momentum components of the wavefuntion, with values $+p\bs{\hat{y}}$ and $-p\bs{\hat{y}}$, as the one for a free particle with momentum $+p\bs{\hat{y}}$, since the spin measurement is made when the particle has momentum $+p\bs{\hat{y}}$ and the particle spin should not change with the reflection on the potential barrier. Analogously, if the spin measurement is made while the particle propagates in the $-\bs{\hat{y}}$ direction, the Wigner rotation is  the one of a free particle with momentum $-p\bs{\hat{y}}$ for both momentum components. If, on the other hand, the particle is confined in a potential well with both momentum components $+p\bs{\hat{y}}$ and $-p\bs{\hat{y}}$ while the spin measurement is made, the quantization axis is given by the average field seen by the particle, such that no Wigner rotation occurs with the change of reference frame. 

In the three examples of the particle state preparation described in the previous paragraph, the Wigner rotation is the same for both momentum components of the state of Eq. (\ref{psi}). After tracing out spin, the spatial wavefunction is pure, such that in the new frame the probability density of finding the particle in different positions is given by Eq. (\ref{p0}). So we conclude that the linear application of the Wigner rotations to the quantum state of Eq. (\ref{psi}) is not a valid procedure, since it lead us to the probability density of Eq. (\ref{p1}) in the new frame. This example demonstrates that we cannot in general linearly apply the Wigner rotations to the quantum state of a relativistic particle without considering the appropriate physical interpretation.

\section{Wigner rotations and the detection of genuine free particle states}

\begin{figure}\begin{center}
  \includegraphics[width=7cm]{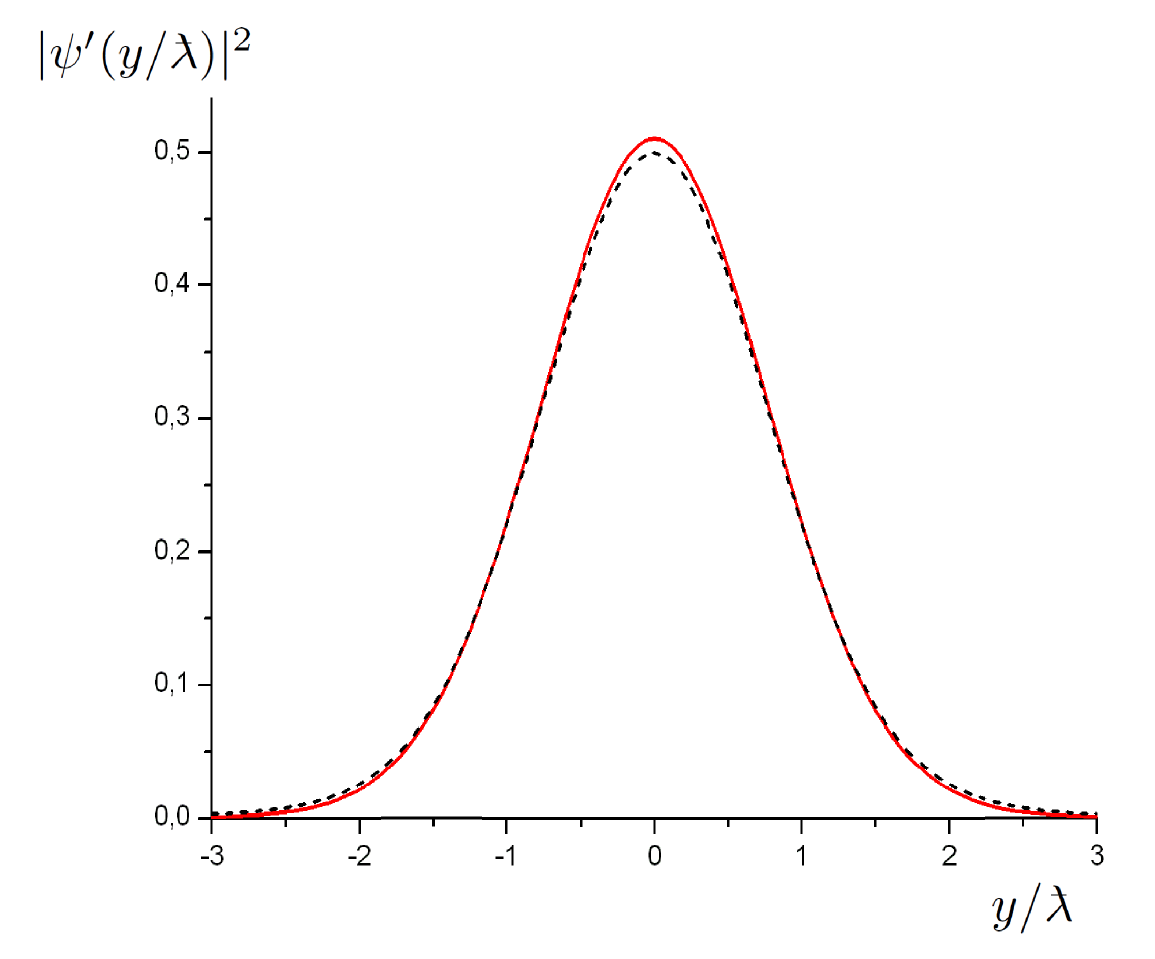}\\
  \caption{Modulus squared of the position wavefunction of a particle in a quantum state with a momentum wavefunction $\tilde{\psi}(p)\propto \mathrm{e}^{-p^2/(2m^2c^2)}$ and spin state pointing in the $z$ direction in reference frame  $S^{(0)}$ (continuous red curve) and in a reference frame $S^{(1)}$ moving with velocity $0.995c\bs{\hat{z}}$ (dashed black curve).
}\label{fig2}
 \end{center}\end{figure}

Before concluding, we would like to briefly discuss the influence of the Wigner rotations on the detection of genuine free particles states in different reference frames. A particle in a superposition of momenta $\tilde{\psi}(p)\propto \mathrm{e}^{-p^2W^2/2}$ in the $y$ direction in the reference frame $S^{(0)}$, for instance, is what we call a genuine free particle state, since the state can be constructed without reflections of the wavefunction, with a uniform potential in the whole space. We will consider here that the spatial wavefunction $\psi(y)$ is given by Eq. (\ref{wavefunction}) with $K(p^0)\propto\sqrt{m/p^0}$ \cite{grandy}. In Fig. \ref{fig2} we plot $|\psi'(y)|^2$ in reference frame  $S^{(0)}$ and $|\psi'(y)|^2$ in a reference frame  $S^{(1)}$ that moves with velocity $\beta \bs{\hat{z}}$ in relation to $S^{(0)}$ when the particle spin in $S^{(0)}$ is prepared in an eigenstate of $\hat{\sigma}_z$ for $W=\lambdabar/\hbar=1/(mc)$ and $\beta=0.995c$. The difference between the two cases is very small, specially when it is considered that the detection probability corresponds to the integral of the square modulus of the wavefunction in regions greater than $\lambdabar$. This difference does not increase much by choosing other values for $W$ and $\beta$.  However, the probabilities of finding the particle in each region must be exactly the same in both frames. So, if in some cases the difference is found to be above the quantum fluctuations, this indicates that the definition of the wavefunction and/or detection probability used are non-physical.

\section{Conclusion}

To summarize, we have shown that the linear application of the momentum-dependent Wigner rotations to the quantum state of a massive relativistic particle in a superposition of counter-propagating momentum states in combination with a general model for particle detection leads to a paradox, since the probability of finding the particle at different positions would depend on the reference frame. Considering the physical implementation of the quantum state, we discussed that the Wigner rotation depends on the preparation method, such that, with a change of the reference frame, the spin transformation of a state in a superposition of different momenta is not necessarily equivalent to the linear application of the momentum-dependent Wigner rotation to each momentum component of the state, a conclusion that solves the paradox. The present work, together with our previous works on the subject \cite{saldanha12a,saldanha12b}, show that relativistic quantum transformations cannot in general be computed only by following a mathematical procedure. The physical meaning of the transformations must always take precedence.

It is worth to mention that it may be possible that by modeling the particle detection by some more complicated scheme the paradox could be solved keeping the linearity of the Wigner rotations. But note that the position operator would have a very complicated dependence on the particle momenta and spin in this case. Although we do not rule out such possibility, we believe that the solution we present for the paradox is more reasonable due to its simplicity and clear physical interpretation in the relativistic quantum information context.

We acknowledge M. Lanzagorta and T. Crowder \cite{lanzagorta17} and E. R. F. Taillebois and A. T. Avelar \cite{taillebois} for calling our attention to the inadequacy of the example presented in the previous version of this work \cite{saldanha13} due to a mistake in the calculations.
P.L.S. was supported by the Brazilian agencies CAPES, CNPq and FACEPE. V.V. acknowledges financial support from the Templeton Foundation, the National Research Foundation and Ministry of Education in Singapore and the support of Wolfson College Oxford.


\end{document}